\documentclass[floatfix,twoside, twocolumn,aps]{revtex4}
 
\usepackage{epsfig}

\begin{document}

\title{Are volatility correlations in financial markets related to Omori processes occurring on all scales?}

\author{Philipp Weber$^{1,2}$, Fengzhong Wang$^1$, Irena Vodenska-Chitkushev$^1$,\\Shlomo Havlin$^{1,3}$ and H.~Eugene Stanley$^1$}

\address{$^1$Center for Polymer Studies and Department of Physics, Boston University, Boston, MA 02215, USA}
\address{$^2$Institut f\"ur Theoretische Physik, Universit\"at zu K\"oln, 50937 K\"oln, Germany }
\address{$^3$Minerva Center and Department of Physics, Bar-Ilan University, Ramat-Gan 52900, Israel}

\date{\today}

\begin{abstract}

We analyze the memory in volatility by studying volatility return
intervals, defined as the time between two consecutive fluctuations
larger than a given threshold, in time periods following stock market
crashes.  Such an aftercrash period is characterized by the Omori law,
which describes the decay in the rate of aftershocks of a given size
with time $t$ by a power law with exponent close to 1. A shock
followed by such a power law decay in the rate is here called Omori
process. Studying several aftercrash time series, we show that the
Omori law holds not only after significant market crashes, but also
after ``intermediate shocks''. Moreover, we find self-similar features
in the volatility. Specifically, within the aftercrash period there
are smaller shocks that themselves constitute Omori processes on
smaller scales, similar to the Omori process after the large crash. We
call these smaller shocks subcrashes, which are followed by their own
aftershocks.  We also find similar Omori processes after intermediate
crashes in time regimes without a large market crash. By appropriate
detrending we remove the influence of the crashes and subcrashes from
the data, and find that this procedure significantly reduces the
memory in the records.  Our results are consistent with the hypothesis
that the memory in volatility is related to Omori processes present on
different time scales.

\end{abstract}

\maketitle

The correlations of stock returns are important for risk estimation, and
can be used for forecasting financial time series.  The absolute value
of the return, which is a measure for volatility, seems to have a memory
\cite{Wood85, Harris86, Admati88, Schwert89, Chan91, Bollerslev92,
Gallant92, Baron92, Ding93, Dacorogna93, Pagan96, Granger96, Liu97,
Cont98, Pasquini99, Liu99, Plerou2001}, so that a return is more likely
to be followed by a return with similar absolute value, which leads to
periods of large volatility and other periods of small volatility
(called volatility clustering in economics). While the absolute value
exhibits long-term correlations decaying like a power law~\cite{note1}, the
correlations of the return itself decay exponentially with a
characteristic time scale of 4 minutes \cite{Liu97, Liu99}.

Recent studies \cite{Yam+05, Wang+06, Vodenska06, Wang++06} reveal
more information about the temporal structure of the volatility time
series by analyzing volatility return intervals, the time between two
consecutive events with volatilities larger than a given
threshold. These return intervals display memory and volatility
clustering, and also scaling properties for different thresholds,
which seem to be universal for different time scales and
markets~\cite{Yam+05, Wang+06, Vodenska06, Wang++06}. This behavior is
similar to what is found in earthquakes \cite{Livina05} and climate
\cite{Bunde04, Bunde05}.  Rare extreme events like market crashes constitute
a substantial risk for investors, but these rare events do not provide
enough data for reliable statistical analysis. Due to the scaling
properties, it is possible to analyze the statistics of return
intervals for different thresholds by studying only the behavior of
small fluctuations occurring very frequently, which have good
statistics.

Lillo and Mantegna found that after a major stock market crash the
rate of volatilities larger than a given threshold $q$ decreases like
a power law with an exponent close to 1~\cite{Li+03}. This behavior is
analogous to the classic Omori law describing the aftershocks
following a large earthquake~\cite{Omori1894}.

Here, we show that the Omori law holds not only after significant
market crashes, but also after ``intermediate shocks''. Moreover, we
find self-similar features in the volatility. Specifically, within the
aftercrash period (characterized by the Omori law) there are smaller
shocks that themselves behave like the Omori law on smaller scales. We
call these shocks subcrashes, which can be considered as ``new crashes
on a smaller scale'', followed by their own aftershocks.

Furthermore, we analyze the memory in volatility return intervals
after large market crashes, and show that the memory is related to the
Omori law.  Indeed, if we perform appropriate detrending, the return
intervals show significantly less memory, but some memory still
exists, independent of the large market crash. We also show that at
least part of this ``remaining memory'' can be described by the
self-similar subcrashes: if we remove also Omori processes due to
subcrashes, the memory is further reduced. We also analyze the memory
in the volatility time series and show that removing the influence of
the major crash and some of its subcrashes reduces the memory in the
dataset. However, some memory still remains so that these crashes
cannot account for the entire memory, raising the possibility that the
``remaining memory'' is due to other subcrashes whose influence was
not removed.

This paper is organized as follows. Section I presents information
about the analyzed data. In section II we show and discuss the
mechanism based on Omori processes on different scales. In section III
we study the memory in return intervals induced by large and
intermediate shocks. In section IV we analyze the influence of crashes
on the volatility memory, and section V presents discussion and
conclusions.

\section{The data sets analyzed}

In order to capture a variety of market crashes, we analyze three
different data sets.

\begin{itemize}

\item{(i)} We study the 1 minute return time series
of the S\&P500 index from 1984 to 1989.  We analyze the aftercrash
period in the 15,000 trading minutes (approximately two months) after
``Black Monday'', 19 October 1987, as well as after a smaller crash on
11 September 1986. We also analyze the time after several other
smaller market crashes within the entire data set.

\item{(ii)} The second data set consists of the TAQ data base of the year
1997 which is provided by the NYSE and contains all trades and quotes
for all stocks traded at NYSE, NASDAQ, and AMEX. We choose the 100
most frequently traded stocks and calculate an index by a summation of
the normalized prices of each stock (normalized by the price). From
this index, we calculate a 1 minute return time series for our
analysis, which we analyze in the approximately two months after the
crash on 27 October 1997.

\item{(iii)} As an example of a crash that is clearly due to an external
event, we also study the 1 minute return series of General Electric
(GE) stock in the three months after 11 September 2001.

\end{itemize}

For all three data sets, we calculate the volatility as the absolute
value of the 1 minute return, normalized by the standard deviation
$\sigma$ of the entire period. Hence, in this paper the volatility and
also the threshold $q$ are measured in units of the standard
deviation $\sigma$.

%
\begin{figure}
  \centerline{ \epsfig{file=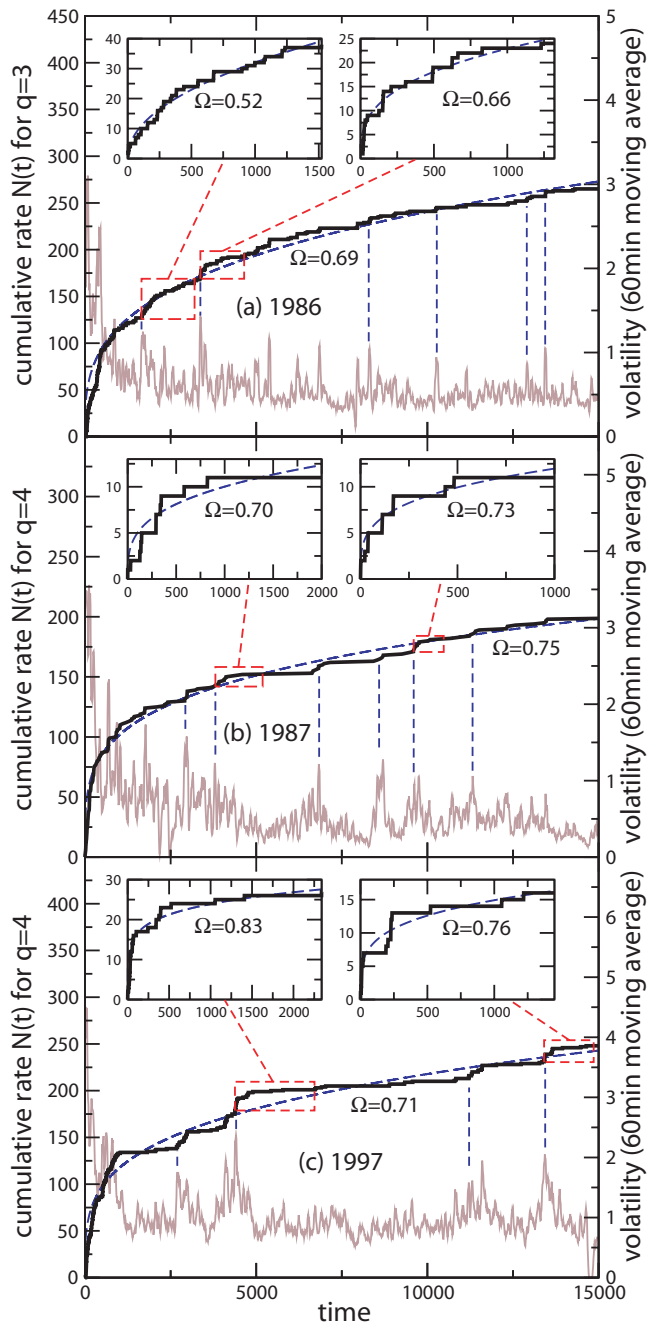,width=8.5cm}}

  \caption{(Color online) Comparison between volatility and the
  cumulative rate $N(t)$ of volatilities (absolute 1 minute returns)
  larger than a threshold $q$. The plots show the 15,000 minutes
  (approximately two months) after the market crashes on (a) 11
  September 1986, with $q=3$, (b) 19 October 1987, with $q=4$, and (c)
  27 October 27 1997, with $q=4$. The volatility is displayed as a
  moving average over 60 minutes in order to suppress insignificant
  fluctuations. The insets show the self-similarity of the data set
  meaning that while the big crash in the beginning introduces a
  behavior following the Omori law, some of the aftershocks introduce
  again a similar behavior on a smaller scale.}

  \label{omori_selfsimilar.fig}
\end{figure}
%

\section{Omori law on different scales}

Lillo and Mantegna \cite{Li+03} showed that the Omori law
\cite{Omori1894} for earthquakes also holds after crashes of large
magnitude in financial markets, so that the rate $n(t)$ of events with
volatility larger than a given threshold $q$ decays as a power law
\begin{equation}
n(t) = k t^{-\Omega} \ \ ,
\label{omori.eq}
\end{equation}
where $\Omega$ is around 1 for large $q$ and $k$ is a parameter
characterizing the amplitude of the rate $n(t)$. For estimating the
parameter $k$ and the exponent $\Omega$, we use the cumulative number
$N(t)$ of events larger than $q$, given by
\begin{equation}
N(t) = \int_0^t n(t')dt' = k \frac{1}{1-\Omega} t^{1-\Omega} \ \ .
\label{omori_cumulative.eq}
\end{equation}
We study the Omori law on different time scales.
Fig.~\ref{omori_selfsimilar.fig} shows the cumulative rate $N(t)$
above (a) $q=3$ and (b,c) $q=4$ compared to the volatility
in time periods following three significant market crashes in (a)
1986, (b) 1987, and (c) 1997.  The volatility is smoothed by a moving
average over 60 minutes in order to remove insignificant
fluctuations. The large shock in the beginning of the time interval is
followed by aftershocks, which induces an Omori-like behaviour of
$N(t)$ (Omori process), shown by the dashed lines representing a power
law fit. However, as seen in Fig.~\ref{omori_selfsimilar.fig} (see
insets) many of these aftershocks seem to behave like ``real'' crashes
with their own aftershocks (subcrashes), but on a smaller scale (shown
by vertical lines). The insets show that a closer look into many of
these subcrashes reveals a similar pattern as the Omori law on large
scales. The exponent $\Omega$ is often smaller after smaller crashes, which
is consistent with the finding that the power law decay of the
volatility after smaller shocks has a smaller exponent than after
large crashes \cite{Sornette03}.  Below we explore the possibility
that the self-similarity of the volatility (where the Omori law is
present on different scales) is directly related to the memory.

\section{Return interval memory after crashes and subcrashes}

In order to explore the memory effects of the Omori law, we first
analyze time periods after very large market crashes. Specifically, we
study the memory in the volatility return intervals, which form a
sequence of time intervals $\tau(t)$ between two consecutive events
with volatilities larger than a given threshold $q$ \cite{Yam+05,
Wang+06, Vodenska06, Wang++06}. We next show that the influence of the
Omori law on $\tau(t)$ can be estimated by comparing the original
$\tau(t)$ with a detrended time series $\tilde{\tau}(t)$ which is
independent of the market crash. We fit the cumulative rate $N(t)$ in
the period after a market crash with a power law according to
Eq.~(\ref{omori_cumulative.eq}), thus obtaining the parameter $k$ and
the exponent $\Omega$ for the rate $n(t)$ \cite{Li+03}. Using $n(t)$, we
can detrend the return interval time series $\tau(t)$ by rescaling by
$n(t)$~\cite{Corral04}
\begin{equation}
\tilde{\tau}(t) = \tau(t) n(t) \ \ .
\label{scaled_tau.eq}
\end{equation}
The rational for this detrending is the following: immediately after
the crash we have a large rate $n(t)$ of high volatilities so that the
return intervals $\tau(t)$ are very short. Later, the rate of high
volatilites becomes small while the return intervals get
large. According to Eq.~(\ref{scaled_tau.eq}), high (low) rates and
small (large) return intervals cancel each other so that
$\tilde{\tau}(t)$ is detrended and thus independent of the existence
of the crash, since the trend caused by the crash is no longer present.

%
\begin{figure}
  \centerline{ \epsfig{file=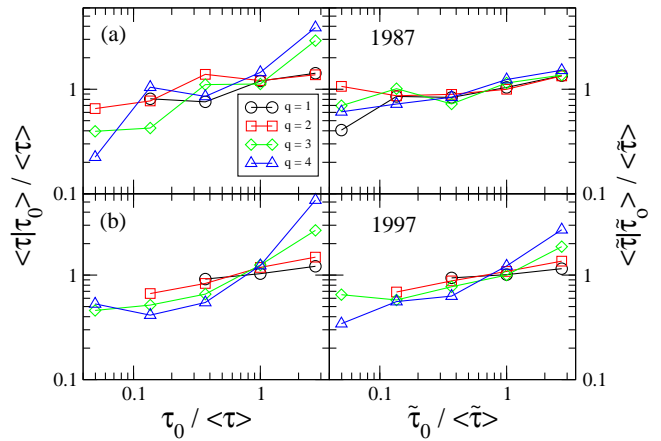,width=8.5cm}}

  \caption{(Color online) Memory in volatility return intervals for
  different thresholds before (left column) and after (right column)
  detrending the time series according to Eq.~(\ref{scaled_tau.eq}).
  The analysis is shown for (a) the S\&P500 index in the two months
  after the crash on 19 October 1987 and (b) an index calculated from
  the 100 most frequently traded stocks from the TAQ data base after
  the crash of 27 October 1997. Removing the Omori law reduces the
  memory in the data sets, but some memory still exists.}

 \label{memory_tau0.fig}
\end{figure}
%

The relation between the Omori law and the short-term memory in the
return interval time series can be studied by analyzing the
conditional expectation value $\left <\tau(t)|{\tau_0} \right>$ of the
return interval series $\tau(t)$ conditioned on the previous return
interval $\tau_0$~\cite{Yam+05, Wang+06}, for both the original return
intervals $\tau(t)$ and the detrended time series $\tilde{\tau}(t)$.
In Fig.~\ref{memory_tau0.fig} (left column), $\left <\tau(t)|{\tau_0}
\right>$ is plotted against $\tau_0$. Both quantities are normalized
by the average return interval $\left<\tau\right>$, for return
intervals after the crashes in (a) October 1987 and (b) October
1997. The deviations from a horizontal line at 1 for all thresholds
show memory: large (small) values of $\tau_0$ are more likely to be
followed by large (small) values of $\tau(t)$. The slopes of the
curves for the detrended time series $\tilde{\tau}$ are significantly
less steep (right column), indicating that detrending the Omori law
from the time series significantly reduces the memory, but some of the
memory still remains, which might be due to the Omori process still
present on smaller scales (see Fig.~\ref{omori_selfsimilar.fig}).

%
\begin{figure}
  \centerline{ \epsfig{file=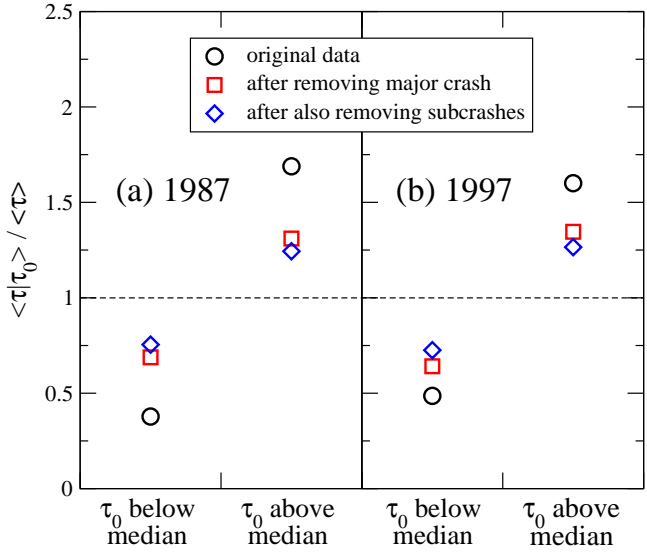,width=8.5cm}}

  \caption{(Color online) Memory in volatility return intervals for
  threshold $q=3$ for (a) the S\&P500 index in the two months
  after the crash on 19 October 1987 and (b) for an index calculated
  from the 100 most frequently traded stocks from the TAQ data base
  after the crash of 27 October 1997. The conditional expectation
  value $\left<\tau|\tau_0\right>/\left<\tau\right>$ conditioned on
  the previous return interval $\tau_0$ is smaller than 1 if $\tau_0$
  is below the median while
  $\left<\tau|\tau_0\right>/\left<\tau\right> > 1$ if $\tau_0$ is
  above the median, indicating the memory in the records
  (circles). The effect gradually weakens upon detrending the time
  series by removing the influence of the major crash (squares) and
  even further when removing some subcrashes (diamonds).}

 \label{memory_tau0_2bins.fig}
\end{figure}
%

In addition to the effect of the major crash, we can also analyze the
influence of Omori processes after subcrashes on smaller scales. To
this end, we further detrend the time series by removing some
subcrashes and test whether the memory is further reduced. After
identifying the subcrashes,
\cite{note2}
we detrend the return intervals $\tau(t)$ by removing the Omori
process due to the major crash as well as the Omori processes induced
by the subcrashes. To this end, we estimate the parameters $k$ and $\Omega$
in Eq.~(\ref{omori.eq}) for the rate $n(t)$ after the major crash as
well as for the rate $n_s(t)$ in the 1000 minutes following each
subcrash (or the time to the next subcrash, if smaller). Note that
$n_s(t)$ is calculated from the detrended return intervals
$\tilde{\tau}(t)$. Then, the double detrended return interval time
series is given by
\begin{equation}
\tilde{\tilde{\tau}}(t) = \left\{ \begin{array}{ccc}
n_s(t)\tilde{\tau}(t) & \mbox{in time following a subcrash}\\
\tilde{\tau}(t) & \mbox{otherwise.}
\end{array}\right.
\label{tau_subcrashes.eq}
\end{equation}
In order to improve the statistics for testing the effect of removing
also subcrashes on the memory, we plot in
Fig.~\ref{memory_tau0_2bins.fig} the conditional expectation value
$\left<\tau|\tau_0\right>/\left<\tau\right>$ for only two $\tau_0$
intervals: $\tau_0$ below and $\tau_0$ above the median of $\tau$.  We
see in Fig.~\ref{memory_tau0_2bins.fig} that when $\tau_0$ is below
the median, $\left<\tau|\tau_0\right>/\left<\tau\right> < 1$, while
$\left<\tau|\tau_0\right>/\left<\tau\right> > 1$ for $\tau_0$ above
the median. This indicates the memory in the records, and also shows
that the memory in the original records (circles) gradually weakens
upon detrending the time series by removing the influence of the major
crash (squares) and further weakens when also some subcrashes are
removed (diamonds). Hence, not only a large market crash but also
smaller subcrashes contribute to the memory in return intervals.

%
\begin{figure}
  \centerline{ \epsfig{file=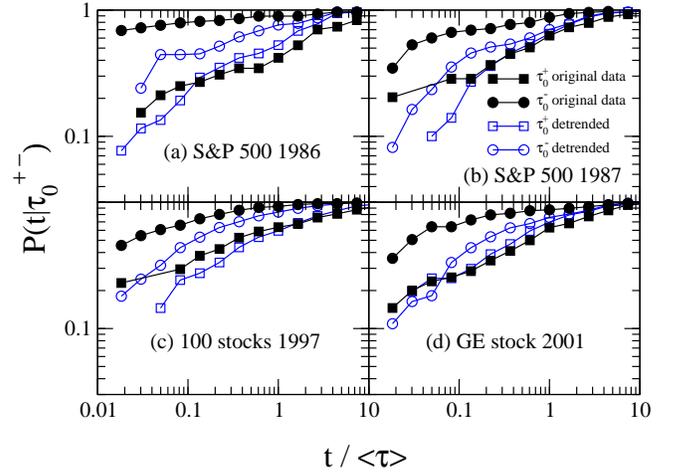,width=8.5cm}}

  \caption{(Color online) Probability $P(t|\tau_0)$ that after a
  return interval $\tau_0$ the next volatility larger than a threshold
  $q=4$ ($q=3$ in (d)) occurs within time $t$.  Here, $\tau_0$ belongs
  to either the 25\% smallest values ($\tau_0^{-}$, circles) or the
  25\% largest values ($\tau_0^{+}$, squares) of $\tau$. The memory in
  the original time series (filled symbols) is reduced by detrending
  according to Eq.~(\ref{scaled_tau.eq}) (open symbols), but some of
  the memory still remains. The results are shown for (a) the S\&P500
  index after a crash on 11 September 1986, (b) the S\&P500 index
  after the crash on 19 October 1987, (c) an index created from the
  100 most frequently traded stocks from the TAQ database after the
  crash on 27 October 1997 and (d) GE stock after 11 September 2001.}

  \label{memory_time.fig}
\end{figure}
%

To further investigate the effect of removing the memory induced by
aftershocks, we analyze the probability $P(t)$ that after an event
larger than a certain volatility $q$ the next volatility larger than
$q$ appears within a time $t$ \cite{Bunde05, Livina05, Vodenska06}.
In order to study the memory, we plot the conditional probability
$P(t|\tau_0)$ for different values of the preceding return interval
$\tau_0$. Fig.~\ref{memory_time.fig} shows $P(t|\tau_0)$ for
$q=2$ under the condition that the preceding return interval
$\tau_0^-$ belongs to the the smallest 25\% of the return intervals or
that the preceding return interval $\tau_0^+$ belongs to the largest
25\%.  The memory in the time series leads to a splitting of the
curves because after larger return intervals (squares) the time to the
next volatility above $q$ is usually large, while it is short after
small return intervals (circles). After detrending the time series the
curves get closer, indicating a reduced memory, but also here some
memory still remains.

%
\begin{figure}
  \centerline{ \epsfig{file=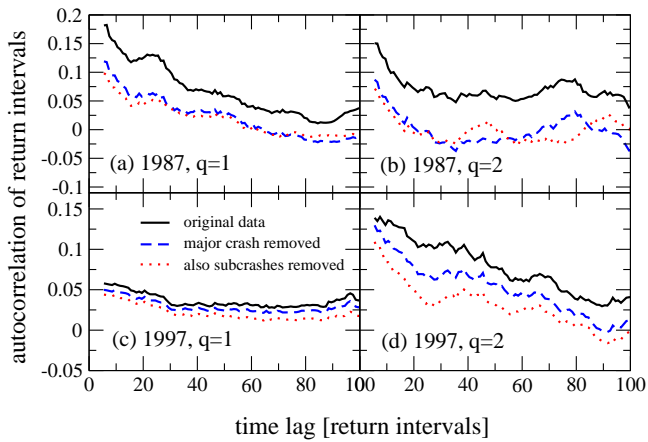,width=8.5cm}}

  \caption{(Color online) Autocorrelation function of the return
  interval time series for threshold (a,c) $q=1$ and (b,d)
  $q=2$. The first row (a,b) shows results from the S\&P500 index in
  the three months after the market crash on October 19, 1987, while
  the second row (c,d) results from an index created from the 100 most
  frequently traded stocks from the TAQ database after the crash on 27
  October 1997.  The Omori law due to the market crash (original data,
  solid lines) induces correlations leading to an offset in the
  autocorrelation function which is removed in the detrended
  $\tilde{\tau}$ (dashed lines), but the data still shows some
  long-term correlations even after removing the influence of the
  Omori law. However, after further detrending with respect to some
  subcrashes (dotted line), the autocorrelation is further
  reduced. All lines are smoothed by a moving average over 10 return
  intervals.}

  \label{autocorr.fig}
\end{figure}
%

To test the long-term memory effects of the Omori process on the
volatility return intervals we study the autocorrelation function
shown in Fig.~\ref{autocorr.fig} for return intervals after the market
crashes in 1987 and 1997 for two different thresholds $q=1$ and
$q=2$. For both thresholds, we see that there exists a significant
correlation even between return intervals 100 steps apart, which
corresponds to approximately 2 to 5 days in 1987 (0.5 to 2 days in
1997) since the average return intervals are
$\left<\tau(q=1)\right>=6.33$ min and $\left<\tau(q=2)\right>=17.4$
min in 1987 and $\left<\tau(q=1)\right>=2.47$ min and
$\left<\tau(q=2)\right>=7.66$ min in 1997. If we now remove the effect
of the Omori process due to the market crash by detrending according
to Eq.~(\ref{scaled_tau.eq}), the memory in the detrended sequence
$\tilde{\tau}$ is reduced significantly, as we see in the dashed
curves of Fig.~\ref{autocorr.fig}. The dotted lines show that removing
also the influence of some subcrashes according to
Eq.~(\ref{tau_subcrashes.eq}) further reduces the memory.

\begin{figure}
  \centerline{ \epsfig{file=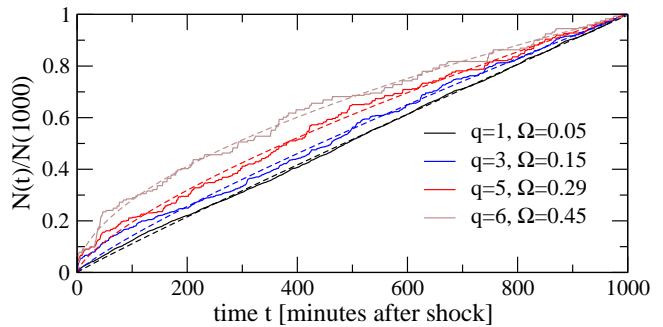,width=8.5cm}}

  \caption{(Color online) Cumulative rate $N(t)$ of events larger than
  a threshold $q$ averaged over the 1000 minutes after 22 shocks
  between $11\sigma$ and $16\sigma$ in the S\&P500 one minute time
  series of the years 1984 to 1989. The data for each shock is
  normalized by $N(1000)$ in order to make different shocks comparable
  irrespective of the current trading activity.  The cumulative rate
  can be well fitted by a power law according to
  Eq.~(\ref{omori_cumulative.eq}). The exponent grows from $\Omega=0.05$ to
  $\Omega=0.45$ for $q=1 \dots 6$.}

  \label{omori_fit_smallshocks.fig}
\end{figure}

So far, we showed indications that within the time period after a big
crash there might exist smaller crashes that behave in a similar
way. The question arises whether such subcrashes are only typical
after a big crash or whether they appear in all time periods
independent of the existence of a big crash. To test this, we analyze
if Omori processes exist also for smaller crashes. We study 22 crashes
of sizes between 11 and 16 standard deviations in the S\&P500 time
series from 1984 to 1989. These crashes are considerably smaller than
the huge crashes of more than 30 standard deviations in a 1 minute
interval studied above. We analyze the cumulative rate $N(t)$ in the
1000 trading minutes following these smaller crashes.  In order to
make different crashes comparable irrespective of the current trading
activity, we normalize the cumulative rate $N(t)$ by $N(1000)$.
Fig.~\ref{omori_fit_smallshocks.fig} shows this normalized rate
$N(t)/N(1000)$ averaged over all aftershock periods~\cite{note3}. For different thresholds $q$, $N(t)/N(1000)$ can be fit
with a power law, Eq.~(\ref{omori_cumulative.eq}). The exponent $\Omega$
increases with the threshold, but is generally smaller than the
exponents found after very large shocks. Our results for the {\it
rate} decay are analogous to volatility studies \cite{Sornette03,
Zawadowski04} where the exponent characterizing the {\it volatility}
decay depends on the magnitude of the shock \cite{Sornette03}.  These
results indicate that relatively small crashes have similar Omori
processes which may lead to memory effects.

%
%

\section{Memory in Volatility after crashes and subcrashes}

In the previous sections, we showed that the memory in return
intervals decreases when we remove effects due to Omori
processes. Since the studied return intervals $\tau(t)$ are derived
from the volatility time series $v(t)$, it would be interesting to
test whether the memory in $v(t)$ is also affected by Omori
processes. Thus, we next analyze the memory in the volatility time
series directly. It is known that a market crash induces a power
law decay of the approximate form
\begin{equation}
v_{PL}(t) \equiv v_0 t^{-\beta}
\label{vola_power.eq}
\end{equation}
with an exponent $\beta \approx 0.2-0.3$ \cite{Li+03, Sornette03}.
In order to study the memory induced by this decay, we compare the
original time series $v(t)$ to a detrended one
\begin{equation}
\tilde{v}(t) \equiv \frac{v(t)}{v_{PL}(t)}
\label{vola_detrended.eq}
\end{equation}
so that $\tilde{v}(t)$ does not depend on the market crash.

\begin{figure}
  \centerline{ \epsfig{file=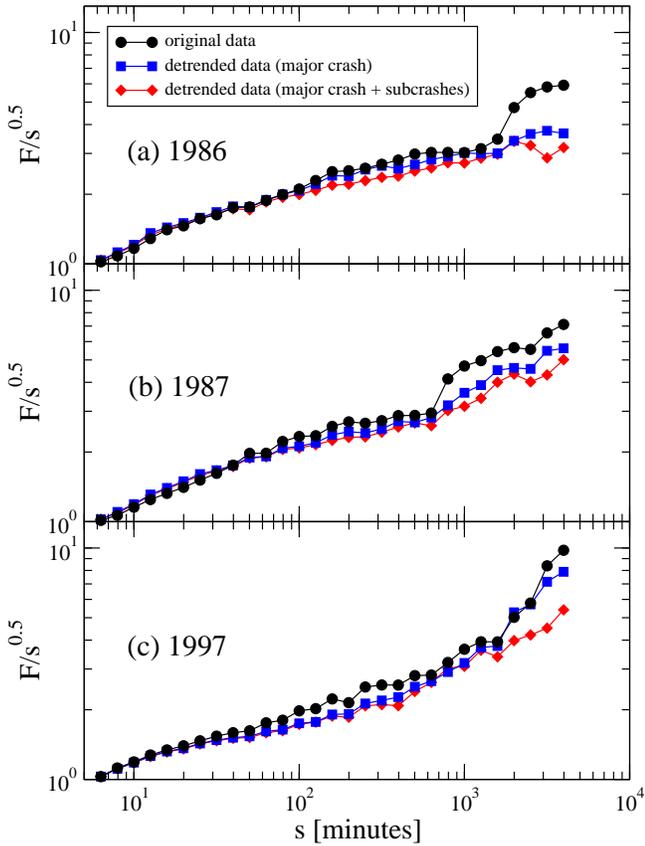,width=8.5cm}}

  \caption{(Color online) Root mean square fluctuation $F(s)$ obtained
  by the second order DFA method (DFA2) for the volatility in the
  15000 minutes following market crashes in (a) the S\&P500 index on
  11 September 1986 and (b) on 19 October 1987, as well as (c) the
  market crash on 27 October 1997 for an index created from TAQ-data
  for 100 stocks. $F(s)$ is divided by $s^{0.5}$ to clarify the
  deviation from uncorrelated data.  Compared to the original
  volatility $v(t)$ (circles), the memory is reduced in the detrended
  records $\tilde{v}(t)$ (squares), and even further after also
  detrending some subcrashes in $\tilde{\tilde{v}}(t)$ (diamonds).}

  \label{dfa_volatility.fig}
\end{figure}

We use second order detrended fluctuation analysis
(DFA2)~\cite{Peng94, Bunde00} to study the long-term memory in the
volatility~\cite{Liu99, Wood85, Harris86, Admati88, Schwert89, Chan91,
Bollerslev92, Gallant92, Baron92, Ding93, Dacorogna93, Pagan96,
Granger96, Liu97, Cont98, Pasquini99, Plerou2001}. In DFA2, the
deviations $F(s)$ (root mean square fluctuations) from a second degree
polynomial fit of the profile
\begin{equation}
y(t)=\sum_{t^\prime=0}^{t} v(t^\prime)
\label{profile.eq}
\end{equation}
as a function of different scales $s$ (time windows) reveal
information about the memory. If $F(s)\sim s^{\alpha}$, the
autocorrelation exponent $\gamma$ of the time series is related to the
exponent $\alpha$ by $\alpha = 1 - \gamma/2$. For $\alpha > 0.5$, the
time series is long-range correlated, it is anti-correlated for
$\alpha < 0.5$, and $\alpha = 0.5$ indicates no long-range
correlations.  Fig.~\ref{dfa_volatility.fig} shows
$\log(F(s)/s^{0.5})$ plotted against $\log s$ for 15,000 trading
minutes after three different market crashes of 1986, 1987, and
1997. With no long-term correlations, the function would be constant,
while a positive slope indicates long-term correlations. For all
crashes, the original time series (circles) shows an increased slope
on large time scales. After detrending according to
Eq.~(\ref{vola_detrended.eq}) and replacing $v(t^\prime)$ by
$\tilde{v}(t^\prime)$ in Eq.~(\ref{profile.eq}), the curve (squares)
gets less steep, indicating a reduction of the memory (the curves are
shifted so that they start at the same point).
%
%

As described before, there are also subcrashes which may induce their
own power law decay on a smaller scale -- not only in the rate, but
also in the volatility. In order to analyze the memory due to these
subcrashes, we further detrend the time series and test whether the
memory is reduced even further. To this end, we fit the detrended
volatility $\tilde{v}(t)$ in the 1000 minutes following each subcrash
(or the time to the next subcrash, if shorter) with a power law
$\tilde{v}_{PL}$ according to Eq.~(\ref{vola_power.eq}). Then, we
further detrend $\tilde{v}(t)$ in these regions using
Eq.~(\ref{vola_detrended.eq}) for $\tilde{v}(t)$ instead of $v(t)$.
The DFA2 curve for the double detrended time series
$\tilde{\tilde{v}}(t) \equiv \tilde{v}/\tilde{v}_{PL}$ is shown in
Fig.~\ref{dfa_volatility.fig}.  The decrease in the slope shows that
the memory is further reduced after removing the influence of the
subcrashes. However, we clearly see that removing the trends induced
by a market crash as well as subcrashes only slightly reduces the
memory in the volatility on quite small scales ($s < 60$ min).

%
\begin{figure}
  \centerline{ \epsfig{file=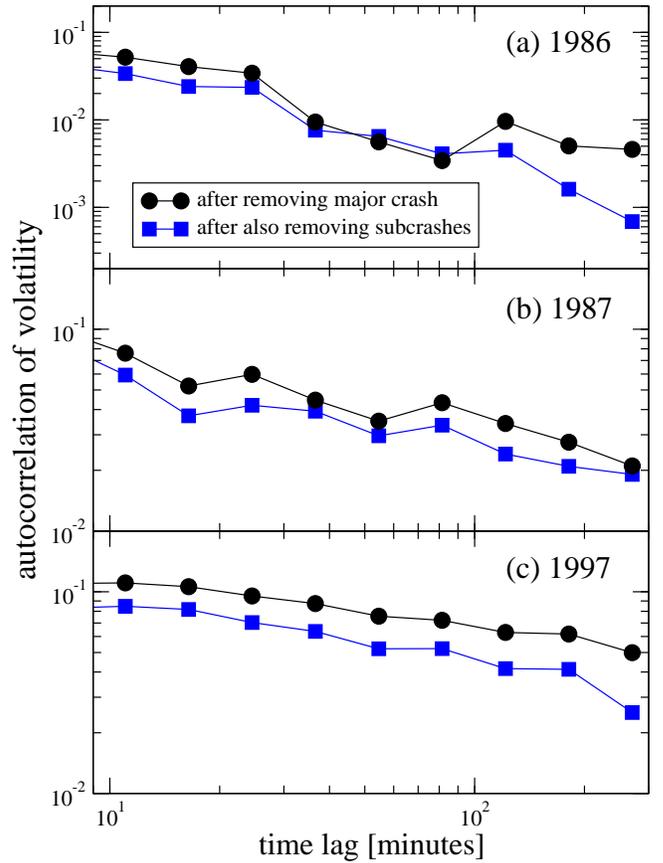,width=8.5cm}}

  \caption{(Color online) Autocorrelation function of the volatility
  time series after detrending. Compared to the volatility time series
  after detrending the major crash (circles), detrending subcrashes
  (squares) further reduces the autocorrelations. The results are
  shown for (a) the S\&P500 index after a crash on 11 September 1986,
  (b) the S\&P500 index after the crash on 19 October 1987, (c) an
  index created from the 100 most frequently traded stocks from the
  TAQ database after the crash on 27 October 1997.  The
  autocorrelation function of the original volatility time series is
  not shown because it is not meaningful as it is dominated by the
  influence of the market crash.}

  \label{autocorr_volatility.fig}
\end{figure}
%

The effect of removing subcrashes on the long-term correlations of
volatility is seen better in Fig.~\ref{autocorr_volatility.fig}. Here,
we compare the autocorrelation functions of the detrended volatility
$\tilde{v}(t)$ and the double detrended volatility
$\tilde{\tilde{v}}(t)$ after also removing subcrashes. It is seen that
generally the autocorrelation of $\tilde{\tilde{v}}(t)$ is smaller
than of $\tilde{v}(t)$, which indicates that the Omori processes after
subcrashes also contain some memory.

\section{Discussion and conclusions}

We find that the volatility exhibits some self-similar features,
meaning that Omori processes exist not only on very large scales, but
a similar behavior is also induced by less significant shocks in the
aftershocks.  After a large market crash, some of the aftershocks can
be considered as subcrashes that initiate Omori processes on a smaller
scale.

We ask the question whether this self-similarity can be responsible
for the memory in volatility return intervals as well as in the memory
of the volatility itself. Our results show that a significant amount
of memory is induced by these crashes and subcrashes, which suggests
that a large part of the memory in volatility might be due to Omori
processes on different scales.

\section*{Acknowledgments}

We thank D. Fu, X. Gabaix, P. Gopikrishnan, V. Plerou, J. Nagler,
B. Rosenow, F. Pammolli, A. Bunde, and L. Muchnik for collaboration on
aspects of this research, and the NSF and Merck Foundation for
financial support.

\newpage


\begin{thebibliography}{99}
\bibitem{note2}
To properly identify subcrashes that can be removed from the records,
we filter the time series with an appropriate criteria for each data
set. For the S\&P500 index time series, including the crashes from
1986 and 1987, we define a subcrash as an event where the 60 minute
moving average of the 1 minute volatility exceeds 1 standard deviation
(corresponding to a much larger 1 minute volatility burst). We also
require at least 500 minutes to the next subcrash (events within 100
minutes are considered as the same subcrash). For the data from 1997,
we analyze the 10 minute moving average, and a subcrash has to exceed
2.5 standard deviations. The other parameters are the same as for the
S\&P500 data.
\bibitem{note3} The average only includes crashes where the volatility
exceeds the threshold $q$ at least 5 times during the studied time
period of 1000 minutes. For e.g.~$q=6$, there are 11 crashes that
satisfy this criteria.
\end{thebibliography}
\end{document}